\RequirePackage{fix-cm}
\documentclass[twocolumn,epjc3]{svjour3}  
\smartqed  % flush right qed marks, e.g. at end of proof
\RequirePackage{graphicx}
\usepackage{amssymb}
\usepackage{textcomp}
\usepackage{graphicx}
\usepackage{epsfig}
\usepackage{amsmath}
\usepackage{slashed}
\usepackage{units}
\usepackage{hyperref}
\usepackage{array}
\usepackage{lmodern}
\usepackage{gensymb}
\usepackage{float}
\usepackage{amsmath}
\RequirePackage[numbers,sort&compress]{natbib}

\journalname{Eur. Phys. J. C}

\begin{document}

\title{Probing neutron-hidden neutron transitions with the MURMUR experiment}

\author{Coraline Stasser\thanksref{e1,addr1}
\and
Guy Terwagne\thanksref{addr1}
\and
Jacob Lamblin\thanksref{addr2}
\and
Olivier M\'{e}plan\thanksref{addr2}
\and
Guillaume Pignol\thanksref{addr2}
\and
Bernard Coup\'{e}\thanksref{addr3}
\and
Silva Kalcheva\thanksref{addr3}
\and 
Steven Van Dyck\thanksref{addr3}
\and 
Micha\"{e}l Sarrazin\thanksref{e2,addr4,addr1}
}

\institute{LARN, University
of Namur, 61 rue de Bruxelles, B-5000 Namur, Belgium \label{addr1}
\and 
LPSC, University of Grenoble-Alpes, CNRS/IN2P3, 53 Avenue des Martyrs,
F-38026 Grenoble, France \label{addr2}
\and 
SCK$\cdot$CEN, Boeretang 200, B-2400 Mol,
Belgium \label{addr3}
\and
Institut UTINAM, CNRS/INSU, UMR 6213,
University of Bourgogne-Franche-Comt\'{e}, 16 route de Gray, F-25030 Besan\c con Cedex,
France \label{addr4}
}

\thankstext{e1}{coraline.stasser@unamur.be (corresponding author)}
\thankstext{e2}{michael.sarrazin@ac-besancon.fr (corresponding author)}

\date{Received: date / Accepted: date}

\maketitle

\begin{abstract}
MURMUR is a new passing-through-walls neutron experiment designed to constrain neutron-hidden neutron transitions allowed in the context of braneworld scenarios or mirror matter models. A nuclear reactor can act as a source of hidden neutrons, such that neutrons travel through a hidden world or sector. Hidden neutrons can propagate out of the nuclear core and far beyond the biological shielding. However, hidden neutrons can weakly interact with usual matter, making possible for their detection in the context of low-noise measurements. In the present work, the novelty rests on a better background discrimination and the use of a mass of a material - here lead - able to enhance regeneration of hidden neutrons into visible ones to improve detection. The input of this new setup is studied using both modelizations and experiments, thanks to tests currently performed with the experiment at the BR2 research nuclear reactor (SCK$\cdot$CEN, Mol, Belgium). A new limit on the neutron swapping probability $p$ has been derived thanks to the measurements taken during the BR2 Cycle 02/2019A: $p<4.0\times 10^{-10} \;  \text{at 95\% CL}$. This constraint is better than the bound from the previous passing-through-wall neutron experiment made at ILL in 2015, despite BR2 is less efficient to generate hidden neutrons by a factor of $7.4$, thus raising the interest of such experiment using regenerating materials.
%\keywords{Matter disappearance–reappearance
%\and Hidden sector \and Mirror matter \and Braneworlds \and Neutron}
\end{abstract}

\section{Introduction}
In many extensions of both the Standard Model (SM) of particle physics and the $\Lambda CDM$ cosmological model, the visible sector of particles containing the usual baryonic matter is supplemented by a "hidden" sector. These models aim at addressing the shortcomings of the standard models, in particular the lack of a microscopic description of Dark Matter, or they aim at exploring the possible low-energy phenomenology associated with quantum gravity. A hidden sector could take the form of a duplication of the SM particle content, in which case each particle would have a hidden mirror partner \cite{art39, art66, art67, art34, art68, art69, art70, art71, art32}. Alternatively, a hidden sector could take a geometrical meaning in the braneworld hypothesis: if our visible world is confined on a brane inside a bulk of more than 4 spacetime dimensions, the bulk could also contain a parallel brane where hidden particles could travel \cite{art26, art31, art28, art74, art29, art60, art75, art76, art73, art65, art4, art38, art77, art30}. 

Here, we consider the generic subclass of hidden sector models for which the fermions could exist in a hidden state as well as in the visible state. In particular the neutron $n$ would have a hidden state $n'$. The hidden state is sterile as it does not interact via the SM interactions with the visible matter. However, the hidden state could mix with the visible state through mass mixing (in the theoretical context of mirror matter \cite{art39, art32, art64}), or through kinetic or geometric mixing (in the context of the parallel brane model \cite{art65, art4, art38, art5, art6, art24}). The mixing induces $n \to n'$ (or $n' \to n$) transitions. The low energy phenomenology (see \cite{art65, art4, art38} for a derivation in the context of the parallel brane model) is obtained by considering the evolution of the four-state quantum problem with the following Hamiltonian:

\begin{equation}
H=\begin{pmatrix}
E_{v} & &  \epsilon \\
\\
\epsilon^{*} & & E_{h}
\end{pmatrix}.  \label{GeneralHiddenSector}
\end{equation}

Here, $E_v$ and $E_h$ describe the energy of the visible and hidden states, respectively, and the mixing is described by $\epsilon$. Those are in principle $2\times 2$ Hermitian matrices acting on the spin states, which we assume diagonal to simplify the following discussion without loss of generality. In the regime of quasi-degenerate states, i.e. $|E_v-E_h |< \epsilon$, neutrons would undergo slow oscillations to the hidden state with large amplitude. Several measurements searching for resonant disappearance of ultracold neutrons have been planed \cite{art64, art44, art37} or reported \cite{art45, art46, art48, art47, art63, art49}, setting already stringent limits in the quasi-degenerate regime. On the contrary, in the regime of non-degenerate states, i.e.  $|E_v-E_h | \gg \epsilon$, the probability for a visible neutron to be found in the hidden state oscillates rapidly, with a mean value of \cite{art38, art5}:

\begin{equation}
p = \frac{2\ \epsilon^2} {(E_{v}-E_{h})^2} \label{Generalp}
\end{equation}

The non-degenerate regime can be probed through regeneration effects, i.e. $n \to n' \to n$ transitions. First, a visible neutron would convert into a hidden neutron by scattering on nuclei in a converter (in practice the moderator material of a large neutron source). Indeed, a collision with a visible nucleus acts as a quantum measurement process and the neutron wave-function collapses either in the visible state with a probability $(1-p)$ or in the hidden state with a probability $p$. The cross section for the $n \to n'$ transition at each collision is given by \cite{art5}:

\begin{equation}
\sigma(n \to n') = \frac{1} {2} \ p \ \sigma_{s}(n \to n),
\label{sigma}
\end{equation}
where $\sigma_s (n \to n)$ is the usual elastic scattering cross section. After the transition, the hidden neutron would propagate freely and escape the source. Outside the source, hidden neutrons could convert back to visible neutrons by scattering on nuclei in a regenerator, by using the reverse process  $n' \to n$, which has the same weak cross section as the direct process $n \to n'$ given by Eq. (\ref{sigma}). 

Experiments based on $n \to n' \to n$ transitions \cite{art5, art6, art24} are called neutron passing-through-wall experiments \cite{art5, art6}. Some of us (M.S.,G.P.,J.L.,O.M.,G.T.) have conducted the first measurement of this kind at the Institut Laue Langevin (ILL, Grenoble, France) \cite{art5, art6}. No significant regeneration of neutrons was observed, which set the limit $p<4.6 \times 10^{-10}$ ($95 \%$ C.L.) \cite{art6}. The present paper describes a new passing-through-wall experiment performed at the BR2 reactor (SCK$\cdot$CEN, Mol, Belgium). The novel detector, which we call MURMUR, separates the material to regenerate neutrons through $n' \to n$ transition from the material to detect the visible neutrons and is also designed to improve the background rejection. 

The paper is organized as follows. In section 2 we elaborate on the concept of passing-though-walls neutron experiments and recall how to evaluate the expected flux of regenerated neutrons at the detector location. Section 3 describes the BR2 reactor and the evaluation of the hidden neutron flux. Section 4 describes the MURMUR detector. In section 5 we discuss the neutron background and mitigation, and in section 6 we present the results of the first run of measurement.

\section{General framework of a passing-through-walls neutron experiment} \label{framework}
\begin{figure*} [h!]
\begin{center}
\includegraphics[scale=0.3]{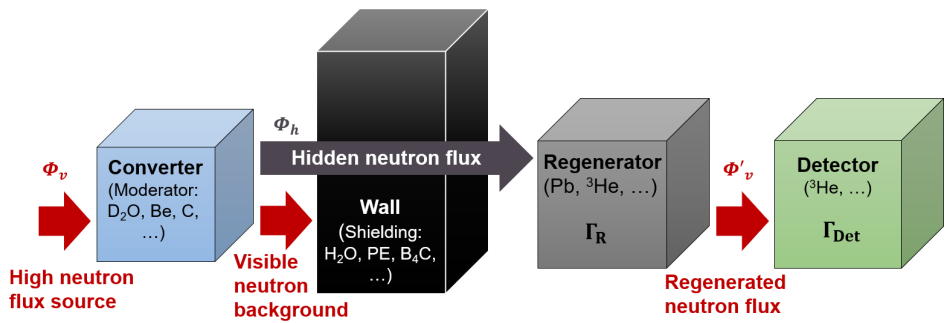} 
\end{center}
\caption{(Color online). Improved concept of a passing-through-wall neutron experiment. Such an experiment is constituted by a converter and a regenerator, both with high scattering cross sections, separated by a shielding wall necessary to shield the rest of the experiment from neutron background. The converter makes possible to transform one part of a visible neutron flux $\Phi_v$ into a hidden one $\Phi_h$, with an efficiency proportional to $p$, while the regenerator enhances their reconversion into visible neutrons, with an efficiency also proportional to $p$, to be detected by a neutron detector separated from the regenerator. $\Gamma_R$ is the regenerated neutron rate in the regenerator. A certain proportion $\xi$ of the regenerated neutrons are detected by the neutron detector with a neutron counting rate $\Gamma_{Det}$.}
\label{Setup}
\end{figure*}

A neutron has a nonzero probability $p$ to convert from a visible state to a hidden one and reciprocally \cite{art5}. It is possible to induce disappearance (reappearance) toward (from) a hidden state thanks to nuclei which possess a high  scattering cross section. The general setup of a passing-through-wall neutron experiment is constituted by a high neutron flux source, a converter, a regenerator and a neutron shielding wall, as shown in Fig. \ref{Setup}. The converter, made of a material with a high scattering cross section, is assumed to generate a flux of hidden neutrons in an hidden state from the incident high neutron flux. Then, a shielding wall makes possible to stop visible neutrons from the source, while hidden neutrons would interact very weakly with the visible matter of the wall and are thus free to pass through it. Finally, a certain proportion of the hidden neutron flux could be regenerated into a visible one thanks to a regenerator, also constituted by a material with a high scattering cross section and with a low absorption cross section. After having swapped back in a visible state, these regenerated neutrons could be detected thanks to a neutron detector, with a neutron detection efficiency $\xi$. 

A nuclear reactor with a high neutron flux, which undergoes many elastic collisions with the moderator of the core, would be then a very intense source of hidden neutrons. In the first experiment of this kind in 2015 \cite{art6}, there was no regenerator distinct from detector. A $33.5$ $\mathrm{cm^3}$ $\mathrm{^3}$He proportional counter at 4 atm simultaneously played the role of regenerator and of neutron detector. In the present new experiment - called MURMUR - the regeneration of hidden neutrons and the detection of regenerated visible neutrons are separated. This improved concept of measurement results into an increase of the efficiency to detect hidden neutrons. Indeed, while the neutron detector remains the same than in the previous experiment, a low-cost lead matrix now surrounds the detector thus enhancing the neutron regeneration as detailed hereafter.

Let us assume that our Universe is endowed with two mutually invisible
sectors. One considers a neutron flux $\Phi _{v}$ inside a
nuclear core. Our aim is to determine the intensity of the hidden neutron
flux $\Phi _{h}$, in the vacuum sector of the core
neighbourhood. For a given volume element of the reactor, since each neutron
collision allows for a conversion into a hidden neutron with the probability $p$, the source of hidden neutrons is proportional to the macroscopic scattering cross section $\Sigma _s$
of the reactor element (first converter in Fig. \ref{Setup}) and to the local neutron flux $\Phi _{v}$. More
specifically, we get the source term corresponding to the number of
hidden neutrons per unit volume and unit of time (see reference \cite{art5}): 
\begin{equation}
S_{h}=\frac 12p\Sigma _s\Phi _{v},  \label{N1}
\end{equation}
Equation (\ref{N1}) is derived by
using a matrix density approach and a two-state Universe description, as shown in ref. \cite{art5}. From
this source term, we deduce the hidden neutron flux $\Phi _{h}$ at the
position $\mathbf{r}$ by integrating over
the reactor volume $V$: 
\begin{equation}
\Phi _{h}(\mathbf{r})=\frac p{8\pi }\int_{}\frac 1{\left| \mathbf{r-r}%
^{\prime }\right| ^2}\Sigma _s(\mathbf{r}^{\prime })\Phi _{v}(\mathbf{r}%
^{\prime })d^3r^{\prime }.  \label{fluxn}
\end{equation}
The relation (\ref{fluxn}) shows that hidden neutron signal should decay as $1/D^2$
when the distance $D$ between the reactor and the detector increases and is valid as neutrons propagate in the vacuum of the hidden sector. $\Phi _{v}$ is derived from ab initio computations using MCNP as explained later. 

We now want to build a detector which could detect the hidden neutron
flux. The present concept is based on the use of a "regenerator" (see Fig. \ref{Setup}) able to convert
the hidden neutron flux $\Phi_h$ into a visible neutron flux $\Phi'_v$ in order to detect it. The calculations of the equations introduced in this section are detailed in \ref{scatt} using an approach similar to this of Ref. \cite{art5}. It can be shown that the visible neutron flux $\Phi'_{v}$ regenerated by the regenerator follows: 
\begin{equation}
D\Delta \Phi'_{v}=-S'_{v}+\Sigma _a\Phi'_{v},  \label{diff}
\end{equation}
where $D$ is now the diffusion coefficient and $\Sigma_a$ the absorption macroscopic cross section of the regenerator medium.

The expression of the regenerated visible source term is given by (see \ref{scatt}):
\begin{equation}
S'_{v}=\frac 12p\Sigma' _s\Phi _{h},  \label{source}
\end{equation}
where $\Phi_h$ is the hidden neutron flux given by equation (\ref{fluxn}) and $\Sigma'_s$ the macroscopic scattering cross section of the regenerator.

Lead has a density much higher than $\mathrm{^3}$He, while having a macroscopic scattering cross section ($\approx 0.4\ \mathrm{cm^{-1}}$) close to the absorption cross section of $\mathrm{^3}$He ($\approx 0.53\ \mathrm{cm^{-1}}$) and a low absorption cross section ($\approx 5.6\times 10^{-3} \ \mathrm{cm^{-1}}$). Thanks to its high atomic mass, lead has also the advantage to thermalize very poorly epithermal and fast neutrons. As a result, lead is an excellent candidate as regenerating medium. However, epithermal and fast neutrons coming from the astrophysical and reactor backgrounds -- although weakly thermalized in lead -- still lead to a residual source of noise for the $\mathrm{^3}$He proportional counter. This is the reason why one deals with the background mitigation in section \ref{LowNoise}.

Moreover, regenerated neutrons can escape out of the lead matrix by scattering. Thus, only part of the regenerated neutrons can be detected by the $\mathrm{^3}$He detector located in the lead matrix (see Fig. \ref{Setup}), resulting then into a regenerated neutron detection efficiency $\xi < 1$. 
Equation (\ref{diff}) can be propagated thanks to MCNP6 numerical computations, as detailed in section \ref{Analysis}, to obtain $\xi$ in order to calculate the regenerated flux inside lead from the measurements of the $\mathrm{^3}$He detector. It is noteworthy that hidden neutrons can also been regenerated directly in the $\mathrm{^3}$He proportional counter \cite{art5}, where a regeneration is here also a detection ($\xi=1$).

\section{The BR2 nuclear core as visible neutron/hidden neutron converter} \label{BR2}
\begin{figure*} 
\begin{center}
\includegraphics[scale=0.25]{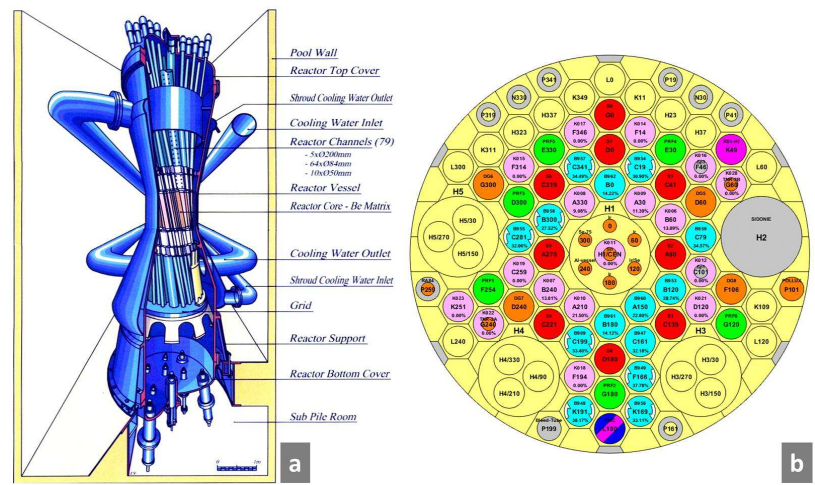} 
\end{center}
\caption{(Color online). (a) The BR2 nuclear reactor geometry of the SCK$\cdot$CEN at Mol in Belgium. (b) Representation of the mid-plan of the BR2 core for the Cycle 02/2019A. Beryllium -- the moderator of the reactor -- appears in yellow, control rods made of hafnium or cadmium are in red and fuel components in pink and cyan. There are $9$ control rods and only the 9th control rod is made of cadmium. It is noteworthy that the core configuration can vary from one run to another. Indeed, the different channels can be filled with fuel, beryllium or samples as required.}
\label{BR22}
\end{figure*}

The Belgian Reactor (BR2) is a highly heterogeneous high flux engineering research thermal reactor operated by the Belgian Nuclear Research Centre (SCK$\cdot$CEN) in Mol, Belgium. This tank-in-pool reactor is cooled by light water in a compact high-enriched uranium (HEU) core (93\% $_{92}^{235}U$). The moderator as well as the reflector of the reactor is a hyperboloid matrix of beryllium surrounding the HEU core (see Fig. \ref{BR22} a). The beryllium matrix is constituted of a large number of skew irregular hexagonal prisms, which form together a twisted hyperbolic bundle around the central $200$ mm channel H1 containing beryllium plug. Due to the particular shape of the beryllium matrix, there is no fixed core configuration, which allows an important flexibility for many different core arrangements. The beryllium matrix holds 79 reactor cylindrical channels: 64 standard channels ($\oslash84$ mm), 10 reflector channels ($\oslash50$ mm) and 5 large channels ($\oslash200$ mm). All channels can contain fuel components, control rods, reflector plugs or experiments. The reactor can be operated at a power of $40-100$ MW, usually about 150 to 180 full power days per year. The thermal neutron flux is about $1\times10^{15}$ $\mathrm{cm^{-2}} \mathrm{s^{-1}}$ and the fast neutron (defined with an energy higher than $0.1$ MeV) flux  $6.0 \times 10^{14}$ $\mathrm{cm^{-2}} \mathrm{s^{-1}}$ at the power of $60$ MW. At the moment, the configuration being used for runs of the experiment is an extended core loaded with about 33 fuel elements and operated at power levels $5$5 MW to $100$ MW with 6 or 8 shim-safety control rods, as seen in Fig. \ref{BR22} b which gives the geometry of the mid-plan of the BR2 for the Cycle 02/2019A. The moderator, i.e the beryllium matrix, appears in yellow and the control rods in red. There are 9 control rods, 8 of which are hafnium and one is cadmium.

\begin{figure} 
\begin{center}
\includegraphics[scale=0.2]{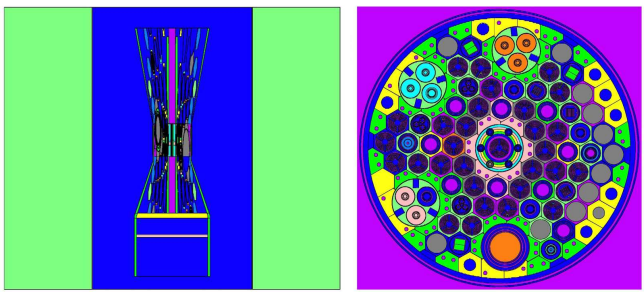} 
\end{center}
\caption{(Color online). Left: MCNP model of the BR2 nuclear core in the YZ-plan. The light water pool appears in blue and the bio-shielding in green. Right: MCNP model of the BR2 mid-plan (XY-plan) for the cycle 02/2019A which included experiments in the core. Cross sections used in the modeling of the core appear in different colors.}
\label{MCNP}
\end{figure}

A 3D model of the BR2 core was developed by the SCK$\cdot$CEN team using the latest version of the Monte Carlo transport code MCNP6 \cite{art21,art22}. The model is a complete 3D description of the BR2 hyperboloid nuclear core formed by twisted and inclined reactor channels (see Fig. \ref{MCNP}). Each channel, fuel component, beryllium plug and experimental device is represented separately, with its own position and inclination. The MCNP6 simulation is coupled with CINDER90 for the isotope evolution part \cite{art21,art22}. The credibility of the MCNP model was demonstrated by multiple comparisons of code predictions with available experimental data, such as control rod worth's, neutron fluxes, gamma heating, linear power and fission rates, reactivity effects, etc \cite{art22}. 

\begin{figure} 
\begin{center}
\includegraphics[scale=0.2]{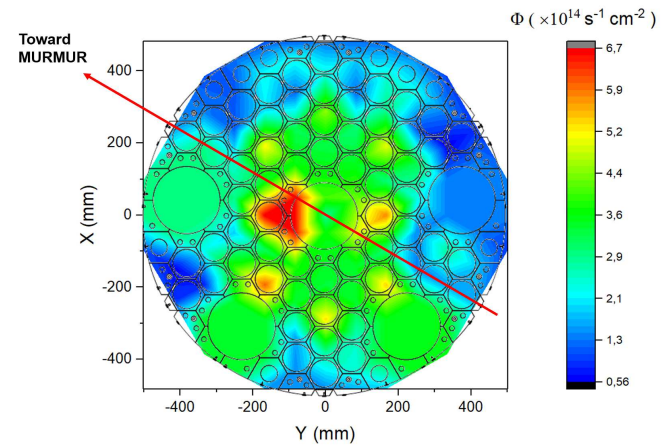} 
\end{center}
\caption{(Color online). Map of the thermal neutron flux in the mid-plan of the BR2 core for the Cycle 02/2019A. Flux has been numerically calculated with MCNP. The direction of the experiment is pointed thanks to a red arrow. The mid-plan of the core is represented in a YX coordinate system, where origin is at the centre of the core. The Z coordinate is perpendicular to the YX plan. }
\label{map}
\end{figure}

The thermal neutron flux (integrated neutron flux for neutron energies below $0.5$ eV) of the Cycle 02/2019A (see the reactor configuration in Fig. \ref{BR22} b) has been numerically calculated at the center of each hexagon of the mid-plan using MCNP model described just above. The results are summarized in Fig. \ref{map}. Thermal neutron flux in a hexagon vary with the Z coordinate. The thermal neutron flux in function of Z is considered constant between $-$20 cm and $+$4 cm and then assumed to drops linearly with Z.

The hidden neutron can be calculated from equation (\ref{fluxn}). As seen in Fig. \ref{BR22} b, the core has a complex geometry. One considers here regular hexagonal prisms as elementary core volumes $dV$, regularly spaced in height and where the thermal neutron flux is constant. One also considers that hexagonal channels are straight and not twisted and inclined (we consider this as negligible since the inclination angle is less than $10\degree$).

The numerical integration is made over the moderator regions of the core, i.e. over the beryllium matrix. Control rod regions, where the flux is the most important, are filled with beryllium in order to be conservative (the insertion of hafnium or cadmium which have a high scattering cross section in the core varies with the elapsed time of the cycle). The rest of the core's volume which is not control rods or the moderator are replaced by aluminum (which has a proper weak scattering cross section). The elementary volumes are calculated with the proper surface in the reference frame in Fig. \ref{map}. The light water pool surrounding the core behaves also as a visible neutrons/hidden neutrons converter. However, the neutron flux at this location being much weaker than in beryllium, its contribution can be neglected.

\section{MURMUR experiment: Lead as a regenerator for hidden neutron} \label{MURMUR}

\begin{figure} 
\begin{center}
\includegraphics[scale=0.22]{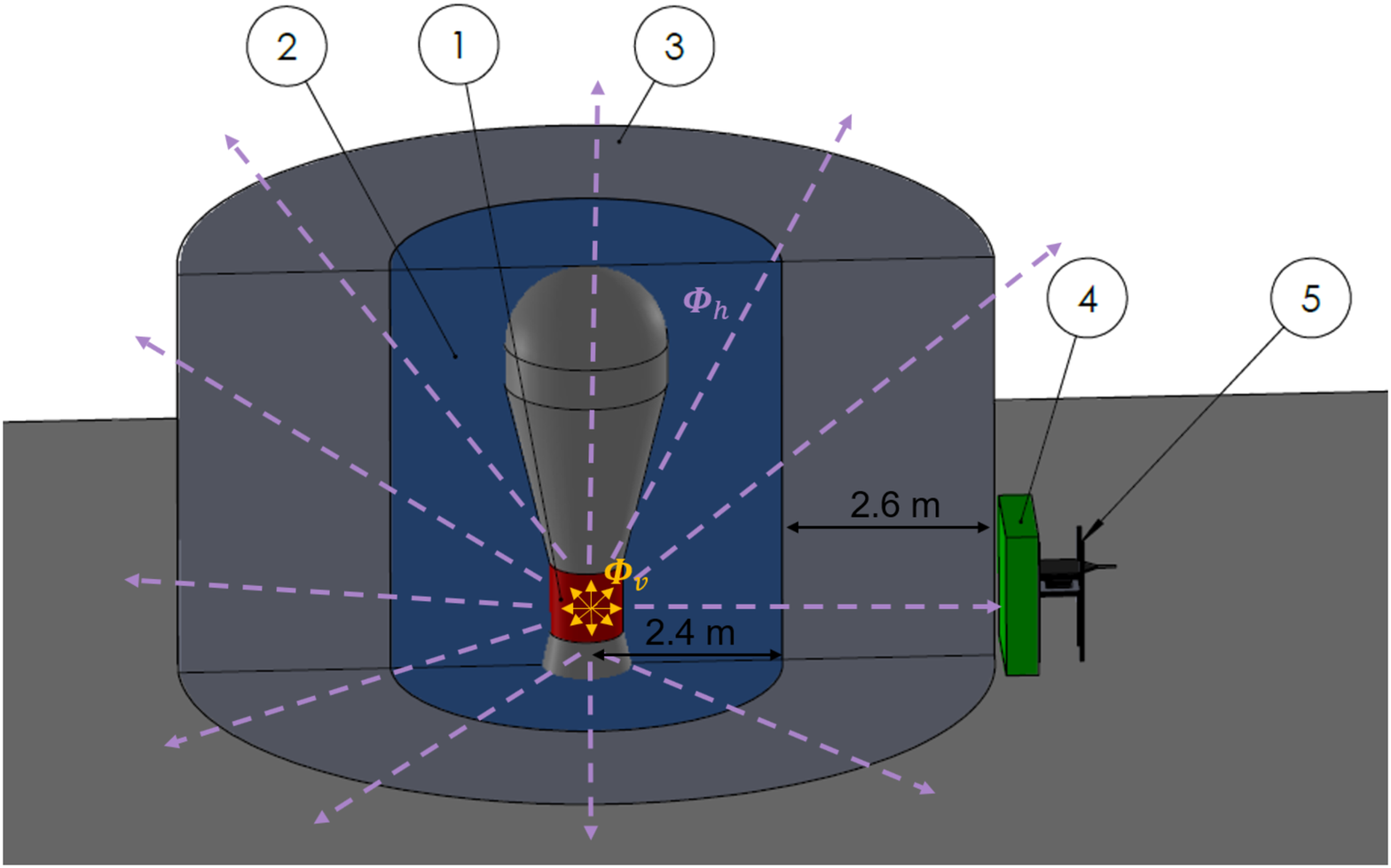} 
\end{center}
\caption{(Color online). Global layout of the MURMUR experiment near the BR2 nuclear reactor of SCK$\cdot$CEN (Mol, Belgium). The BR2 nuclear core (1) provides the visible neutron flux $\Phi_v$. A small part of this visible neutron flux is expected to be converted into a hidden neutron flux $\Phi_h$, which can freely escape the light water pool (2) and the concrete wall (3) of the reactor. In a second configuration of the experiment, a wall made of paraffin (4) has been added in order to reduce the neutron background coming from the nuclear core. In this configuration, the experiment (5) is located at $6.18$ meters from the centre of the BR2 nuclear core. }
\label{3D}
\end{figure}

\begin{figure*} 
\begin{center}
\includegraphics[scale=0.6]{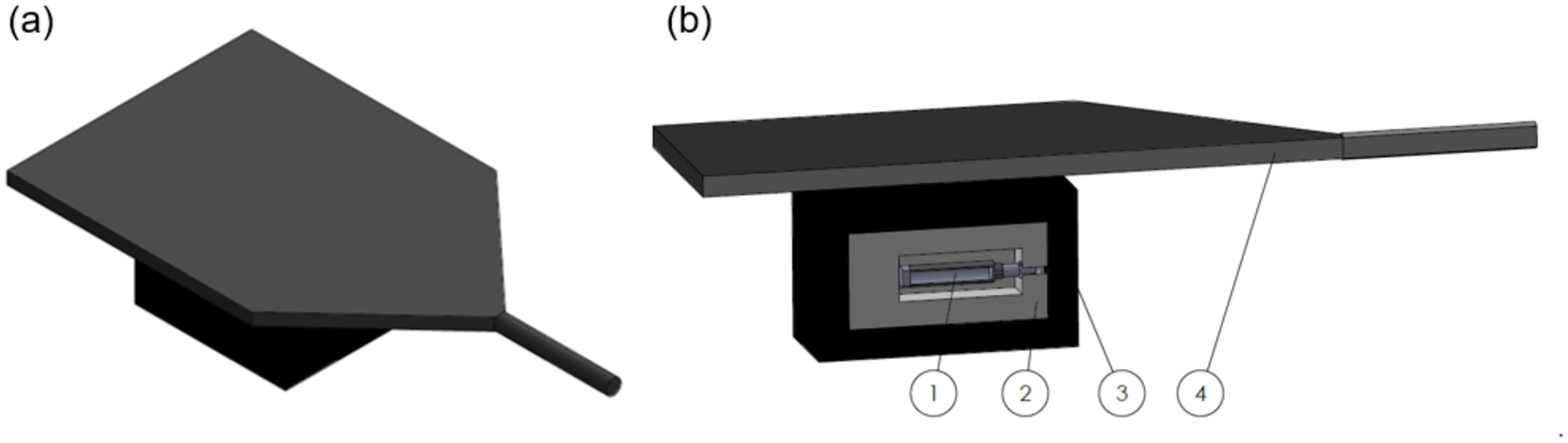} 
\end{center}
\caption{(Color online). Design of the MURMUR detector: (a) complete view, (b) sectional view. A $33.5$ $\mathrm{cm^{3}}$ $\mathrm{^3}$He proportional counter at $4$ atm (1) is embedded in a matrix of $23.6\times 17.7\times 11.5 \ \mathrm{cm^{3}}$ made of $49.82$ kg of lead (2). A $4\pi$ boron carbide box of $3.6\ \mathrm{cm}$ of thickness acts as a passive shielding (3). On the top of the shielding, a $50\times 50 \ \mathrm{cm^{2}}$ EJ200 organic plastic scintillator detects muons and vetos the data acquisition (4).}
\label{Disposition}
\end{figure*}

The MURMUR experiment \cite{art24} is installed at $6.18$ meters from the centre of the BR2 nuclear reactor of the SCK$\cdot$CEN at Mol (Belgium) as shown in Fig. \ref{3D} where the global layout of the experiment near the nuclear core is described. The nuclear core is mainly constituted of beryllium and is surrounded by a light water tank for reactor cooling. A light water pool and an heavy concrete wall constitute the biological shield, which guarantees the security of the BR2 site against radiations. The low-noise MURMUR detector itself is detailed in Fig. \ref{Disposition}.

The beryllium matrix which plays the role of moderator for the BR2 nuclear core acts as a visible neutrons/hidden neutrons converter. The regenerator which regenerates hidden neutrons into visible ones is a $23.6\times 17.7\times 11.5 \ \mathrm{cm^{3}}$ matrix of lead, constituted of lead blocks with a full volume of $4389.4$ $\mathrm{cm^{3}}$ and a total mass of $49.82$ kg. While having a weak absorption cross section, lead has a high macroscopic scattering cross section close to the macroscopic absorption cross section of $\mathrm{^3}$He ($0.53$ $\mathrm{cm^{-1}}$) used for the first experiment at the ILL in 2015 \cite{art5}. Thus, the hidden neutrons can be regenerated into visible neutrons in a large volume of lead, while being weakly absorbed. Regenerated visible neutrons are free to propagate in lead up to an cylindrical $\mathrm{^3}$He proportional counter of $33.5$ $\mathrm{cm^{3}}$  with a gas pressure of $4$ atm located inside the matrix of lead, making possible the detection of hidden neutrons. Hidden neutrons can also be regenerated directly in the $\mathrm{^3}$He counter while being detected at the same time. In addition, due to heavy nuclei, lead thermalizes very poorly neutrons compared with other materials as graphite or heavy water. This property is important, as epithermal or fast neutrons could succeed to pass through the wall shielding and would thermalize in the regenerator, representing then a possible source of noise in the $\mathrm{^3}$He proportional counter. All the regenerated neutrons in lead are not detected by the $\mathrm{^3}$He proportional counter, only a certain proportion $\xi$ of them are captured by the $\mathrm{^3}$He gas. This proportion is numerically calculated in section \ref{Analysis} thanks to MCNP6. A $4\pi$ boron carbide ($B_4 C$) box of $3.6$ cm of thickness is used as shielding wall in order to isolate the lead and the $\mathrm{^3}$He proportional counter from the external environment. A paraffin shielding of $40$ cm of thickness has also been added between the core concrete wall and the experiment in order to slow down epithermal and fast neutrons coming from the nuclear core. A plane plastic organic scintillator EJ200 of $50\times 50$ $\mathrm{cm^{2}}$  is finally placed at the top of the setup in order to veto the data acquisition. Indeed, as discussed in section \ref{LowNoise}, cosmic muons can induce neutrons in lead, becoming a source of noise. The presence of the scintillator on the top of the matrix of lead allows both to monitor the cosmic particle flux and to prevent the data acquisition from the muon-induced neutrons. 

Signals from both detectors of the experiment ($\mathrm{^3}$He counter and scintillator) are connected to a DT5790 digitizer of the CAEN firmware \cite{art10} including a flash ADC (analog-to-digital converter). The acquisition is triggered with a threshold on the signal amplitude and for reach triggered event, the charge is integrated. To allow particle discrimination, two different time windows are used for the integration. Indeed, for a given charge, neutron-captures which induce nuclear recoils (proton and tritium), make shorter tracks than gamma rays which produce electrons, by photoelectric and Compton effects. These differences makes possible to discriminate them according to the shape of their voltage pulses. The Pulse Shape Discrimination ($PSD$) is a number calculated online according to the expression:

\begin{equation}
PSD=\frac{Q_L-Q_S}{Q_L},  \label{PSDeq}
\end{equation}
where $Q_L$ and $Q_S$ are respectively the long and the short integrated charges. Time windows have been chosen empirically in order to optimize the discrimination between $\gamma$ rays and neutrons (see section \ref{PSD}). The long gate is 20 $\mu s$ and the short gate is 1.5 $\mu s$.

The software creates two output files every hour, one for each detector ($\mathrm{^3}$He and scintillator), which contain information about each measured pulse accessible for readout at any time by the user: the long charge, the short charge, the $PSD$ and the treatment time of the pulse by the software in ms. An anti-coincidence system is also implemented in the software in order to veto the $\mathrm{^3}$He proportional counter. This veto is used to overcome the muons-induced neutrons in lead, possible source of noise of the experiment. Each time an event is measured in the $\mathrm{^3}$He proportional counter in a time window of $8$ $\mu s$ after an event detection in the scintillator, it is automatically suppressed by the software and does not appear in the list event output file of the $\mathrm{^3}$He detector.

\section{Background mitigation} \label{LowNoise}
The MURMUR detector must reach a very low rate of background: the lower the background, the better the constrain on the swapping probability $p$. This section describes the different possible sources of background and characterizes the efficiency of the experimental setup to deal with them, in the light of preliminary measurements.

\subsection{Background description}
The sources of background can be classify in two categories. The first one is related to the $\mathrm{^3}$He detector in the form of $\gamma$ rays and intern $\alpha$ particles. The second one gathers the visible neutrons coming from the reactor, or be created in the atmosphere, in the reactor building and surrounding materials by spallation of cosmic muons or $(\alpha,n)$ reactions.

As mentioned above, $\gamma$ rays can interact in the detector by photoelectric effect or Compton scattering. Given the path length of electrons, $\gamma$ rays are at low energy. However, a certain proportion of $\gamma$ ray events, for example, events which have a longitudinal track, can overlap in the region of the neutron event energy spectrum, leading to an overestimate of the neutron rate in the $\mathrm{^3}$He detector.

The $\alpha$-particle emission from the actinides present as contaminants in counter materials is a well known issue of low-background experiments \cite{art11, art12}. Their energy can vary from 0 to several MeV and overlaps in the region of the neutron capture products in the energy spectrum of $\mathrm{^3}$He counters. The $\alpha$ signal can thus not be distinguish from a neutron signal. The mean $\alpha$ emission rate depends on the nature of the material and is more important in aluminum ($\sim 6\times 10^{-4}-3\times 10^{-2}\ \mathrm{s^{-1}} $) than in steel ($\sim 10^{-4}\ \mathrm{s^{-1}} $) \cite{art12}. The $\mathrm{^3}$He proportional counter used for the experiment is made in steel but in reason of the very weak counting rate specific of passing-through-wall neutron experiment ($\sim 10^{-4}\ \mathrm{s^{-1}}$), $\alpha$ emissions can be a source for an undesirable background noise. For this reason, measurements of the natural radioactivity of the $\mathrm{^3}$He proportional counter have been carried out with a HPGe detector in a low-noise system made of one ton of lead surrounded by plastic scintillators for the mitigation of the muon background in the HPGe detector \cite{art61, art62}. We did not measure any difference of signal for the main $\gamma$ rays coming from actinides with and without the $\mathrm{^3}$He counter inside the system, which means that the counter contains a very weak quantity of actinides. The intern $\alpha$ background of the counter can thus be neglected.

Cosmogenic neutrons, which are mainly induced by muon spallation, constitute an important source of background for experiment looking for rare events, as hidden neutrons. In the case of MURMUR, cosmogenic neutrons can thus be induced in the surrounding of the experiment \cite{art15,art16} or directly in the matrix of lead \cite{art14}. From the muon-induced neutron yield \cite{art15,art16} and from the muon spectrum  \cite{art17}, one can roughly estimate that $0.5$ neutron per second is produced in $50$ kg of lead. These neutrons having a mean energy of $8$ $MeV$, the probability they would be captured by the $\mathrm{^3}$He is weak. However, some of the lowest-energy neutrons produced by spallation could be thermalized enough in the lead to become thermal neutrons. Finally, all materials surrounding the experiment, as concrete and shielding, have the ability to produce neutrons under the effect of the cosmic muon flux. These fast neutrons could be energetic enough to pass through the shielding, loss enough energy inside it to reach the thermal energy and have good chance to be detected by the $\mathrm{^3}$He proportional counter. 

Finally, $(\alpha,n)$ reactions can also be a source of background due to the presence of the large quantity of concrete as biological shielding \cite{art51, art52, art53}. 

\subsection{Pulse shape discrimination} \label{PSD}

\begin{figure} [h]
\begin{center}
\includegraphics[scale=0.32]{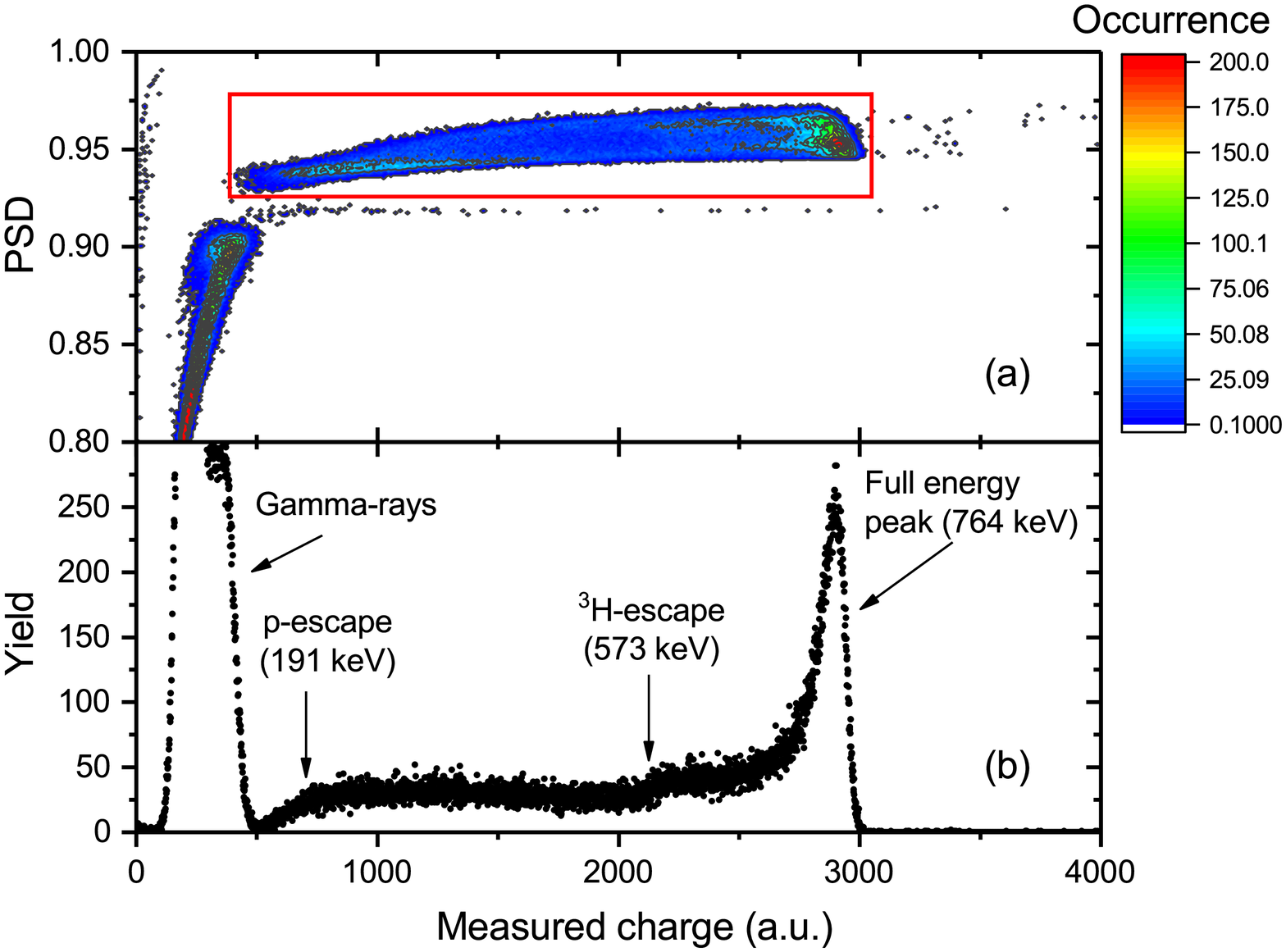} 
\end{center}
\caption{(Color online). (a) $PSD$ quantity as a function of the measured charge in the $\mathrm{^3}$He proportional counter. The discrimination between $\gamma$ rays and neutrons (red box) is very well established. $\gamma$ rays can be entirely filtered with an appropriate $PSD$ cut. (b) $\mathrm{^3}$He counter charge spectrum.}
\label{NeutronPSD}
\end{figure}

Figure \ref{NeutronPSD} shows the $PSD$ as a function of the measured charge from the $\mathrm{^3}$He counter taken for 15 days together with the reconstructed energy spectrum of the $\mathrm{^3}$He counter. It is clearly observed that the neutron spectrum is well reconstructed as we can observe the full energy peak at 764 keV as well as the p and $\mathrm{^3}$H escapes at 191 and 573 keV respectively. We can conclude that the events due to neutrons are clearly discriminated using a $PSD$ selection in addition to the energy cut (see red box in Fig. \ref{NeutronPSD}).

\subsection{Background measurements}

Table \ref{tab1} gives the results of measurements at different locations, at the University of Namur and at the BR2 nuclear reactor in Mol, during run periods
Cycle 05/2018A and Cycle 06/2018A respectively in October and November 2018, and for different configurations: (a) the $\mathrm{^3}$He counter alone inside the boron carbide box, (b) the $\mathrm{^3}$He counter inside the lead block and the boron carbide box, (c) this setup with the veto added and finally, (d) the veto alone in a region of energy containing fast neutrons and muons. At the Mol site, measurements have been done with or without 40 cm of paraffin between the experiment and the reactor.

First, the measurement of the $\mathrm{^3}$He counter alone in the B4C box at the University of Namur, configuration (1-a), is in agreement with the measurement which was done to prepare the ILL experiment \cite{art6}. The addition of lead (1-b) increases the neutron signal in the $\mathrm{^3}$He detector of one order of magnitude due to muon spallation since the same measurement carried out using the active veto (1-c) comes back to the initial value.  Muons coming mostly from the zenith \cite{art17}, one has chosen to put a scintillator on the top only of the boron carbide box. The sides of the experiment are thus not protected by the veto, even though they profit from the umbrella effect of the edges of the scintillator. The veto time window have to be chosen carefully in order to well filter the muon-induced neutrons while avoiding the unnecessary loss of acquisition time.

The same measurements done at the BR2 nuclear reactor during OFF period, configurations (2) and (4), do not give the same results. We explain the difference by the presence of the reactor building with large thickness of concrete in which the muons can interact by spallation. Therefore, we expect less muons, as seen with configurations (2-b) and (4-b) where the rates are due to muon-induced neutrons in the lead, but more neutrons as seen in configuration (4-a). The distribution of arrival direction is also expected to be less vertical. This explains the lower impact of the active veto. While the rates is reduced by a factor of 8 in Namur, it remains almost unchanged at BR2 (configurations (2-c) and (4-c)).

During the ON period, we can notice a higher rate of neutrons in the $\mathrm{^3}$He counter (3-c) and also in the veto (3-d), coming from the reactor. This reactor-induced background justifies the addition of the paraffin block which allows, by thermalizing high energy neutrons, to reduce significantly the counting rates in both detectors (5-c and 5-d),
With this configuration, there is almost no difference between OFF and ON periods.  

One can conclude that the low-noise efficiency of the MURMUR experiment is limited by the cosmic rays background which induces neutrons both in the environment and in the matrix of lead by the sides.

\renewcommand{\arraystretch}{2}
\setlength{\doublerulesep}{0pt}
\begin{table*} 
\caption{Counting rates at different locations, at the University of Namur and at the BR2 nuclear reactor in Mol, during run periods
Cycle 05/2018A and Cycle 06/2018A respectively in October and November 2018, and for different configurations. More details and discussions are given in the text. Uncertainties are calculated at 68\% C.L.}
\begin{center}
\begin{tabular}{llllll} 

\hline & (1) Namur & \multicolumn {2} {c} {\hspace{3.5cm} Mol} \\ \cline{2-2} \cline{3-6}
  & &  \multicolumn {2} {c} {\hspace{-0.8cm} w/o paraffin} &  \multicolumn {2} {c} {\hspace{-0.5cm} w/ paraffin} \\  \cline{3-6} 
  & & (2) OFF & (3) ON & (4) OFF & (5) ON \\
  \hline (a) $\mathrm{^3}$He counter alone + B4C ($\mathrm{\times 10^{-4} \ s^{-1}}$) & $\mathrm{0.5 \pm 0.1}$ & - & - & $\mathrm{1.5 \pm 0.2}$ & - \\
 (b) $\mathrm{^3}$He counter + Pb + B4C ($\mathrm{\times 10^{-4} \ s^{-1}}$) & $\mathrm{4.2 \pm 1.0}$ & $\mathrm{3.5 \pm 0.3}$ & - & $\mathrm{3.3 \pm 0.3}$ & - \\
 (c) $\mathrm{^3}$He counter + Pb + B4C + veto ($\mathrm{\times 10^{-4} \ s^{-1}}$) & $\mathrm{0.5 \pm 0.2}$ & $\mathrm{2.5\pm 0.2}$ & $\mathrm{4.4 \pm 0.4}$ & $\mathrm{3.3 \pm 0.3}$ & $\mathrm{3.3 \pm 0.2}$ \\
  \hline (d) Veto alone ($\mathrm{s^{-1}}$) & - & $\mathrm{22.46 \pm 0.01}$ & $\mathrm{73.06 \pm 0.02}$ & $\mathrm{30.12 \pm 0.01}$ & $\mathrm{31.54 \pm 0.01}$ \\
  \hline
\end{tabular}
\end{center}
\label{tab1}
\end{table*}

\subsection{Background subtraction} \label{noiseSubstraction}
Some sources of background still remain, as muon-induced neutrons which are not filtered by the veto and neutrons coming from $(\alpha,n)$ reactions. In the aim of removing their contributions, reactor ON and reactor OFF measurements will be realized thanks to BR2's shut-down periods. The neutron counting rate obtained during the OFF period of the reactor could be thus subtracted from the neutron counting rate of the reactor ON measurements. From the signal difference, we thus obtain an upper estimation of the neutron flux induced by the reactor, whatever its origin, possibly due to the generation of hidden neutrons followed by their reappearance. This flux makes possible to fix a new constraint on the swapping probability $p$ thanks to numerical computations. Long acquisition times of about 30 days have been done in order to have enough statistic and thus to reduce the uncertainty on the measurements.

\section{Analysis in term of neutron-hidden neutron transitions} \label{Analysis}

This section describes the method used to derive the constraint on the  neutron swapping probability $p$ thanks to the MURMUR experiment measurements. From Eqs. (\ref{fluxn}) and (\ref{source}), the regenerated neutron rate in lead is a function of $p$:

\begin{equation}
\Gamma_R(E)=p^2 \times d_{Pb}(E),
\label{gammaR}
\end{equation}

where 

\begin{equation}
d_{Pb}(E)=\frac{\Sigma_{s,Pb} V_{Pb}}{16 \pi} \int_{}\ \frac {\Sigma _s(\mathbf{r}^{\prime })\ \Phi _{v}(\mathbf{r}%
^{\prime },E)}{\left| \  \mathbf{r-r}%
^{\prime }\right| ^2}d^3r^{\prime}
\label{d}
\end{equation}
is numerically calculated from the visible neutron flux $\Phi_v$ induced in the BR2 nuclear core and calculated in section \ref{BR2}.

Only a certain proportion of hidden neutrons regenerated in lead will be detected by the $\mathrm{^3}$He proportional counter. Monte Carlo simulations are required to study the propagation of regenerated neutrons in lead and evaluate the proportion of them which are captured by the $\mathrm{^3}$He gas. In this purpose, we used MCNP6. The detection efficiency $\xi$ depends on the neutron energy. Indeed, neutron capture cross section decreases with increasing energies. In order to ensure the relevance of the simulations, MCNP6 results have been confronted to experimental data. For this purpose, a calibration experiment has been carried out with an $\mathrm{^3}$He detector and a neutron source of $1.79 (\pm 0.04)\times 10^{3}$ Bq supplemented by a $50\times 40 \times 10\ \mathrm{cm^{3}}$ high density polyethylene block (HDPE) used to thermalize fast neutrons coming from the source. MCNP6 simulations have been performed with the whole setup and the addition of a thermal treatment card for the HDPE. The number of neutronic captures counted with MCNP6 has been compared to the number of detected neutrons after noise subtraction. Table \ref{AmBe} summarizes the results.

\begin{figure} 

\includegraphics[scale=0.38]{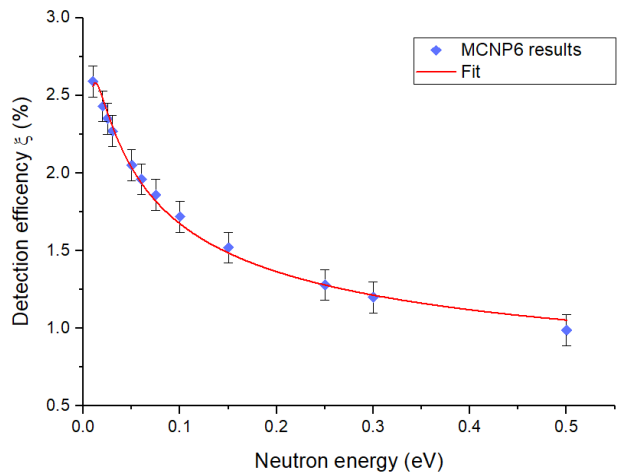} 
\caption{(Color online). Detection efficiency $\xi$ (\%) of the regenerated neutrons in lead by the $\mathrm{^3}$He proportional counter in function of the neutron energy (eV) calculated with MCNP6. Dots are the MCNP6 results and the red line is a fit function of the equation: $A/x+B/\sqrt{x}+C$, where $x$ is the neutron energy, $A=-0.026 \pm 0.002$, $B=0.474 \pm 0.021$ and $C=0.435 \pm 0.048$.}
\label{Fit}

\end{figure}

The detection efficiency of regenerated neutrons has been calculated with MCNP6 against the neutron energy according to the experimental setup geometry. Figure \ref{Fit} shows MCNP6 results for the regenerated neutron detection efficiency $\xi$ against thermal neutron energy. In the thermal energy region (between 0.01 and 0.5 eV), the detection efficiency varies from 2.6 to 1 \%. These values are then convoluted with the thermal energy spectrum of the BR2 \cite{art23} in order to obtain a mean detection efficiency of $1.64\pm 0.17 \%$, where the uncertainty of $0.17\%$ is considered in order to be conservative with table \ref{AmBe}. MCNP6 results have been confirmed thanks to Geant4 numerical computations with a discrepancy between the two software lower than 10 \%.

\setlength{\doublerulesep}{0pt}
\renewcommand{\arraystretch}{1.5}
\begin{table*} [t] 
\caption{Experimental and simulated proportion of neutrons which are detected in the $\mathrm{^3}$He counter while emitted by the AmBe source at 10 and 50 cm from the detector. $R$ is the MCNP/Experimental ratio.}
\begin{center}
\begin{tabular} {llll}

   \hline \textbf{Distance} &  \textbf{Experiment}   &   \textbf{MCNP6}  &   \textbf{R} \\ 
  \hline 10 cm & $\mathrm{1.94 (\pm{0.05}) \times 10^{-3}}$ & $\mathrm{2.28 (\pm{0.55}) \times 10^{-3}}$ &  $\mathrm{1.17 (\pm{0.17})}$ \\
  \hline 50 cm & $\mathrm{8.21 (\pm{2.23}) \times 10^{-5}}$ & $\mathrm{7.54 (\pm{1.70}) \times 10^{-5}}$ &  $\mathrm{0.92 (\pm{0.35})}$ \\
  \hline
\end{tabular}
\end{center}
\label{AmBe}
\end{table*}

\begin{table*}
\caption{Neutron counting rate in the $\mathrm{^3}$He detector with the complete experimental setup and muon/fast neutron counting rate in the veto during the December 2018/January 2019 shutdown period of the BR2 (OFF 1), during the Cycle 02/2019A of April 2019 and during the shutdown period of May/June 2019 (OFF 2). Uncertainties are calculated at 90\% C.L. with a Poisson statistic for neutron and a Gaussian one for veto's events. These measurements are used to derive a new constraint on the swapping probability $p$. The last row gives mean atmospheric pressures during measurement periods calculated thanks to weather reports taken by the weather station of R\'{e}tie, near Mol in Belgium.}
\begin{center}
\begin{tabular}{llll} 

\hline 
  & OFF 1 & ON & OFF 2\\
 \hline
 $\mathrm{^3}$He+Pb+B4C+veto ($\mathrm{\times 10^{-4} \ s^{-1}}$) & $\mathrm{3.09^{+0.17}_{-0.16} }$ & $\mathrm{3.10^{+0.18}_{-0.17} }$ & $\mathrm{2.88^{+0.15}_{-0.15}}$\\
 Veto alone ($\mathrm{s^{-1}}$) & $\mathrm{29.88 \pm 0.02}$ & $\mathrm{31.09 \pm 0.02}$ & $\mathrm{29.38 \pm 0.02}$ \\
 Mean atm pressure (hPa) & 1020.6 & 1016.0 & 1016.4 \\
  \hline
\end{tabular}
\end{center}
\label{tab2}
\end{table*}

Table \ref{tab2} gives the measurements used to derive a new constraint on the swapping probability $p$. 760 hours of measurement were considered during the Cycle 02/2019A of the BR2 nuclear reactor in April 2019. Two shut-down periods were considered to do the noise subtraction, respectively before and after the Cycle 02/2019A, i.e. 831 hours of shut-down in December 2018/January 2019 and 998 hours in May/June 2019. The mean atmospheric pressure during these measurements are closed to each other and the counting rate in the $\mathrm{^3}$He counter during the two shut-down periods are statistically compatible, which makes possible to use the mean value of these two measurements to do the shut-down subtraction. The resulting upper constraint $\Gamma_{Det}$ on the hidden neutron flux regenerated into a visible neutron flux, which could be detected in the $\mathrm{^3}$He proportional counter, is then given by:

\begin{equation}
\Gamma_{Det}< 3.3\times 10^{-5}\ \mathrm{s^{-1}} \;  \text{at 95\% CL}, \label{GammaDet}
\end{equation}

which is better than the value in reference \cite{art6} by a factor of $4$ and where confidence intervals have been calculated according to the method described in reference \cite{art25}. This improvement on $\Gamma_{Det}$ compared with the first experiment carried out at ILL \cite{art5, art6} is explained by a better background mitigation (PSD, ON-OFF measurements) and a better statistic due to a longer acquisition time. The numerical calculation of the regenerated neutron rate $\Gamma_R(E)$ in lead from Eqs. (\ref{gammaR}) and (\ref{d}) leads to the value of $d_{Pb}$:

\begin{equation}
d_{Pb}=7.6\times 10^{15}\ \mathrm{s^{-1}}.\label{GammaS}
\end{equation}

The BR2 visible neutron flux $\Phi_v$ has been calculated as mentioned in section \ref{BR2} and can be seen in Fig. \ref{map}. The numerical calculation method of the hidden neutron flux $\Phi_h$ is described in section \ref{BR2}. The value of the beryllium neutron scattering cross section is 6.2 b, which is a convoluted value according to the thermal neutron energy spectrum of the BR2 \cite{art23} (cross sections for different neutron energies come from END/F VII library). The length of the lead matrix ($24$ cm) is not enough to induce a significant variation of the hidden neutron flux along its propagation pathway. The flux is then calculated at the center of the lead block and considered constant everywhere in lead. The neutron scattering cross section of the lead being constant in the the thermal energy range, it does not require to be averaged over the thermal neutron energy spectrum. 

The regeneration of hidden neutrons directly inside the $\mathrm{^3}$He counter is also considered. In this case, a regeneration is a detection, so $\xi=1$. The numerical calculation of the regenerated hidden neutron in the $\mathrm{^3}$He counter by using a formula similar to Eq. (\ref{d}), but where the macroscopic cross section of lead is replaced by the neutron capture cross section of the $\mathrm{^3}$He ($0.53$ $\mathrm{cm^{-1}}$) and the volume by the $33.5$ $\mathrm{cm^{3}}$ of gas, leads to:

\begin{equation}
d_{^3He}=8.6\times 10^{13}\ \mathrm{s^{-1}}.\label{GammaHe3}
\end{equation}

The constraint on the swapping probability $p$ is then given by:
\begin{equation}
p<\sqrt{\frac{\Gamma_{Det}}{\xi(E) \times d_{Pb}(E)+d_{^3He}}}.
\label{formulep}
\end{equation}

Equations (\ref{GammaDet}), (\ref{GammaS}), (\ref{GammaHe3}), and (\ref{formulep}) finally lead to the following constraint on the neutron swapping probability:
\begin{equation}
p<4.0\times 10^{-10} \;  \text{at 95\% CL.}    \label{p}
\end{equation}

This constraint is slightly better than the one obtained in 2016 ($p<4.6 \times 10^{-10}$) \cite{art6} with the first experiment at the ILL (Grenoble , France) \cite{art5}. It must be underlined that the expected efficiency for the BR2 reactor to produce a hidden-neutron flux is weaker than for the ILL research reactor -- despite a similar source power. Indeed, the $d_{^3He}$ parameter of the ILL experiment was $6.4 \times 10^{14}$ s$^{-1}$, i.e. 7.4 times the value at BR2. This is mainly due to the larger effective volume of the reactor moderator at the ILL, as the moderator acts as a source of hidden neutrons. Roughly speaking, the BR2 would produce a flux of hidden neutrons $7.4$ times weaker than at ILL. Then, while the present value for $\Gamma_{Det}$ is better by a factor of $4$ when compared with the previous experiment at ILL \cite{art6}, the new constraint on $p$ should be worse. Nevertheless, the addition of the lead matrix as a regenerator makes possible to increase the detected regenerated neutron flux by a factor equal to $(V_{Pb} \ \Sigma_{Pb}\ \xi + V_{^3He}\ \Sigma_{^3He})/(V_{^3He}\ \Sigma_{^3He})=2.5$ with respect to the first experiment at the ILL in 2016 \cite{art6}.

While the BR2 nuclear reactor presents a lower ability to produce hidden neutrons, MURMUR makes possible to give a constraint better than the previous one found at the ILL \cite{art6} thanks to a lower counting rate and a regenerator made of lead. This result is thus promising for future low-noise passing-through-walls neutron experiments using regenerating materials as lead. Improvements of the MURMUR experimental setup are in progress -- such as a $4\pi$-active veto in development -- in order to reduce the background and to make possible to get a better constraint on $p$ (the constraint on $p$ is as small as the neutron counting rate is, as expressed by Eq. (\ref{formulep})), while continuing to benefit from the long-term location provided by the BR2. MURMUR is assumed to stay in function at BR2 until 2021, which makes possible to test many other experimental configurations.

Although we are not concerned here with the consequences of the constraint given by Eq. (\ref{p}), it is worth noticing that literature offers some clues about the expected values for $p$.
The swapping probability $p$ is related to the neutron-hidden neutron coupling constant $g$ derived from braneworlds models \cite{art4,art38,art50,art78}. Constraints on the neutron swapping probability $p$ makes possible to rule out a large range of parameters for 5 and 6-dimensional braneworld models at the TeV and the GUT scales \cite{art50,art78}. The largest expected value for $p$ in the context of a 5-dimensional bulk for braneworlds at the reduced Planck energy scale is about $1.7 \times 10^{-11}$ \cite{art50}. This value is expected to be reachable by future passing-through-walls neutron experiments like MURMUR, performed near a more efficient source of hidden neutrons.
For neutron-mirror neutron oscillations, the parameter of interest is $\epsilon$ (see Eq. (\ref{GeneralHiddenSector})) and could be extracted from Eq. (\ref{Generalp}) provided the eigenstates of the energy of the visible and hidden states can be constrained or computed, a task beyond the scope of the present paper.

\section{Conclusions}

In the last five years, some concerns have focused on the realization of passing-through-wall neutron experiments able to constrain the existence of hidden sectors thanks to disappearance/reappearance of neutrons in the surroundings of a nuclear core. MURMUR is a new and improved passing-through-wall neutron experiment, installed for now near the BR2 reactor of SCK$\cdot$CEN at Mol (Belgium). Many improvements with respect to the first passing-through-wall neutron experiment carried out at the ILL (Grenoble, France) in 2015 have been achieved. The major ones are a lower neutron counting rate thanks to a background subtraction, a pulse shape discrimination and the use of $50$ kg of lead as a regenerator for hidden neutrons, separated from the neutron detector. This makes possible to obtain a new constraint on the neutron swapping probability $p$ better than the last one found at the ILL \cite{art6}.

\begin{acknowledgements}
The authors wish to acknowledge, posthumously, Edgar Koonen, former BR2 Deputy Reactor Manager, passed away in 2017. His enthusiasm for our work allowed us to initiate the MURMUR collaboration. C.S., G.T. and M.S. acknowledge Louis Lambotte, Tijani Tabarrant and Alexandre Felten for technical support. The authors are grateful to Wim De Vocht and Nico Segers for their help during the installation of the experiment at the BR2 and for their technical support. This work is supported by the Department of Physics of the University of Namur. C.S. is supported by a FRIA doctoral grant from the Belgian F.R.S.-FNRS. M.S., PI of the MURMUR collaboration, acknowledges the University of Namur to host and support the project. 
\end{acknowledgements}

%_________________________________________________________________________________

\appendix

\section{Lead to regenerate hidden neutrons into visible ones}

\label{scatt}

Let us consider a hidden neutron flux $\Phi_h$ generated by a nuclear reactor. We introduce the calculation of the regenerated visible neutron production rate due to interactions between hidden neutrons and a regenerator medium made of nuclei with a high elastic cross section and a weak absorption. Then, our aim is to find a relation
between the regenerated visible neutron flux $\Phi_{v}$ and the hidden neutron flux $%
\Phi _{h}$. More specifically, we look for a source term: 
\begin{equation}
S_{v}=K\Phi _{h},  \label{B0}
\end{equation}
where the constant $K$ in the source term, which will be assessed in the following.
It can be shown that $K=\frac{1}{2}p\Sigma _{s}$, where $\Sigma _{s}$ is the scattering
macroscopic cross section of the regenerator. To do this, we follow the same procedure than in Ref. \cite{art5}. 

\noindent Neutrons are described through the two-sector Pauli equation in which we neglect
the neutron decay process: 
\begin{equation}
i\hslash \partial _t\Psi =\mathbf{H}\Psi ,  \label{A1}
\end{equation}
such that

\begin{equation}
\mathbf{H}=\begin{pmatrix}
E_{v} & &  \epsilon \\
\epsilon^{*} & & E_{h}
\end{pmatrix},  \label{GeneralHiddenSectorbis}
\end{equation}

with $E_v$ and $E_h$ the energy of the visible and hidden state respectively, and $\epsilon=-i \Omega \hbar$ the mixing between the two states, where $\Omega$ depends on the type of coupling. The behaviour of the neutron flux is
described through the use of the matrix density $\rho $ via the Liouville-Von Neumann equation \cite{art7}, 
\begin{equation}
\frac d{dt}\rho =-i\hbar ^{-1}\left( \mathcal{H}\rho -\rho \mathcal{H}%
^{\dagger }\right) +I_c,  \label{A4}
\end{equation}
where $I_c$ is the collisional integral which describes the collisions
between neutrons and nuclei in the medium and is given by the expression:
\begin{equation}
I_c=nv\int F(\theta )\rho F^{\dagger }(\theta )d\Omega.  \label{A5}
\end{equation}
$v$ is the mean relative velocity between neutrons and molecules and $n$ is
the density of molecules (nuclei) in matter. We have:
\begin{equation}
F(\theta )=\left( 
\begin{array}{cc}
f(\theta ) & 0 \\ 
0 & 0%
\end{array}
\right) ,  \label{A6}
\end{equation}
which describes the scattering of neutrons by the molecules and where $%
\theta $ is a scattering angle. The second diagonal term is equal to zero
since we assume there is no matter in the hidden sector.

\noindent We also define $\mathcal{H=}\mathbf{H}\mathcal{+}\mathbf{C}$, with: 
\begin{equation}
\mathbf{C}=\left( 
\begin{array}{cc}
\frac{-2\pi nv\hslash f(0)}{k} & 0 \\ 
0 & 0%
\end{array}
\right) ,  \label{A7}
\end{equation}
which accounts for the presence of matter in the system and where $k$ is the neutron wave vector. Let us use the definition: 
\begin{equation}
\rho =\left( 
\begin{array}{cc}
\rho _{v} & r+is \\ 
r-is & \rho _{h}%
\end{array}
\right) ,  \label{A8}
\end{equation}
where the $r$ and $s$ terms in the off-diagonal elements of the density matrix are responsible of the quantum coherence which allows the swapping of neutrons in another sector. We also define 
\begin{equation}
w_R=4\pi \frac{nv}k\text{Re}(f(0)),\quad w_I=4\pi \frac{nv}k\text{Im}%
(f(0)),  \label{A9}
\end{equation}
and the optical theorem related to the cross section $\sigma _{tot}$: 
\begin{equation}
\sigma _{tot}=\sigma _E+\sigma _I=\frac{4\pi }k\text{Im}(f(0))=\frac{w_I}{nv}%
,  \label{A10}
\end{equation}
where $\sigma _E$ and $\sigma _I$ are the elastic and inelastic (absorption)
cross sections, with 
\begin{equation}
\sigma _E=\int \left| f(\theta )\right| ^2d\Omega .  \label{A11}
\end{equation}
We then obtain the following system from Eq. (\ref{A4}): 
\begin{eqnarray}
\frac d{dt}\rho _{v}=-2\Omega r-nv\sigma _I\rho _{v},  \label{A12}
\end{eqnarray}
\begin{equation}
\frac d{dt}\rho _{h}=2\Omega r,  \label{A13}
\end{equation}
\begin{equation}
\frac d{dt}r=\left( \eta -\frac{1}{2}w_R\right) s-\frac{1}{2}rw_I+\Omega \left( \rho
_{v}-\rho _{h}\right) ,  \label{A14}
\end{equation}
\begin{equation}
\frac d{dt}s=-\left( \eta -\frac{1}{2}w_R\right) r-\frac{1}{2}sw_I,  \label{A15}
\end{equation}
with $\hbar \eta =E_{v}-E_{h}$. Since we consider an initial hidden neutron flux and its leakage in our visible sector, we have the following initial conditions:

\begin{equation}
\label{Acondiinit}
  \rho_v(t=0)=0,
  \quad \rho_h(t=0)=1.
\end{equation}

Looking at Eq. (\ref{A13}), we see that the
neutron population in the hidden sector mainly decreases due to neutron
leakage into our visible sector. We can consider that $2\Omega r$ must be
very weak and we assume that $\partial _t\rho _{h}\approx 0$, i.e., $\rho
_{h}$\ is almost constant at the regenerator scale. From the weakness of the neutron leakage into our visible sector, an additional condition is naturally coming: $\rho_h  \ll \rho_v$. Finally, considering that $\hbar \eta$ is at least equal to the difference between the gravitational energy of the two sectors, then its value is about $100$ eV at least \cite{art38, art6}. We thus have $\eta \gg w_R$. Then, using Eq. (\ref{A10}),
Eqs. (\ref{A12})-(\ref{A15}) can be recast in the following convenient form: 
\begin{eqnarray}
\frac d{dt}\rho _{v}=-2\Omega r-nv\sigma _I\rho _{v},  \label{B12b}
\end{eqnarray}
\begin{equation}
\rho _{h}\sim 1,  \label{B13b}
\end{equation}
\begin{equation}
\frac d{dt}r=\eta s-(1/2)nv\sigma _{tot}r-\Omega \rho _{h},  \label{B14b}
\end{equation}
\begin{equation}
\frac d{dt}s=-\eta r-(1/2)nv\sigma _{tot}s.  \label{B15b}
\end{equation}
Now, Eqs.(\ref{B14b}) and (\ref{B15b}) constitute a simple non-homogeneous
first-order differential equation that is easy to solve since $\rho
_{h}\approx 1$, i.e., since $\rho _{h}$ is constant. Let us set:

%\begin{widetext}
\begin{equation}
\left( 
\begin{array}{c}
r \\ 
s%
\end{array}
\right) =e^{-(1/2)nv\sigma _{tot}t}\left( 
\begin{array}{cc}
\cos (\eta t) & \sin (\eta t) \\ 
-\sin (\eta t) & \cos (\eta t)%
\end{array}
\right) \left( 
\begin{array}{c}
R \\ 
S%
\end{array}
\right) .  \label{C6}
\end{equation}
If we insert Eq. (\ref{C6}) into Eqs.(\ref{B14b}) and (\ref{B15b}), we
obtain: 
\begin{equation}
\frac d{dt}\left( 
\begin{array}{c}
R \\ 
S%
\end{array}
\right) =e^{(1/2)nv\sigma _{tot}t}\left( 
\begin{array}{cc}
\cos (\eta t) & -\sin (\eta t) \\ 
\sin (\eta t) & \cos (\eta t)%
\end{array}
\right) \left( 
\begin{array}{c}
-\Omega \rho _{h} \\ 
0%
\end{array}
\right) ,  \label{C7}
\end{equation}
from which we deduce that: 
\begin{equation}
\left( 
\begin{array}{c}
R \\ 
S%
\end{array}
\right) =-\Omega \rho _{h}\left( 
\begin{array}{c}
\int_0^t\cos (\eta t^{\prime })e^{(1/2)nv\sigma _{tot}t^{\prime }}dt^{\prime
} \\ 
\int_0^t\sin (\eta t^{\prime })e^{(1/2)nv\sigma _{tot}t^{\prime }}dt^{\prime
}%
\end{array}
\right) .  \label{C8}
\end{equation}
%\end{widetext}

Using Eqs.(\ref{C6}) and (\ref{C8}), Eq. (\ref{B12b}) can be rewritten as: 
\begin{eqnarray}
\frac d{dt}\rho _{v} &=&2\Omega ^2\rho _{h}\int_0^te^{-(1/2)nv\sigma
_{tot}(t-t^{\prime })}\cos (\eta (t-t^{\prime }))dt^{\prime }  \notag \\
&&-nv\sigma _I\rho _{v},  \label{C9}
\end{eqnarray}
where we have set $p=2\Omega ^2/\eta ^2$ \cite{art38, art6}. Let us now consider an initial local neutron flux $\Phi _0=u_0v$
where $u_0$ is the initial neutron density. The neutron fluxes in each sector
are given by $\Phi _{v,h }=\Phi _0\rho _{v,h }$. Then, from Eq. (\ref{C9})
we deduce: 
\begin{eqnarray}
\partial_t\Phi _{v} &=& p\eta ^2\Phi _{h}\int_0^te^{-(1/2)nv\sigma
_{tot}(t-t^{\prime })}\cos (\eta (t-t^{\prime }))dt^{\prime }  \notag \\
&&-nv\sigma_{I}\Phi _{v}  \label{B18bis}
\end{eqnarray}
Since the neutron flux $\Phi _{h}=u_{h}v$, where $u_{h}$ is the local
neutron density in the hidden sector, then: 
\begin{eqnarray}
\partial _tu_{v}&=&p\eta ^2\frac 1v\Phi _{h}\int_0^te^{-(1/2)nv\sigma
_{tot}(t-t^{\prime })}\cos (\eta (t-t^{\prime }))dt^{\prime } \notag \\
&& -n\sigma _I\Phi_{v}  \label{B19}
\end{eqnarray}
Since $u_{h}$ is now local, Eq. (\ref{B19}) must be supplemented by a
divergence term $\mathbf{\nabla \cdot j}_{v}$ to account for the local
behaviour of the neutron current $\mathbf{j}_{v}=u_{v}\mathbf{v}$. We then
deduce the continuity equation for neutrons in the second sector:

\begin{equation}
\mathbf{\nabla \cdot j}_{v}+\partial _tu_{v}=S_{v}-\Sigma _a\Phi _{v},
\label{B20}
\end{equation}
which is the continuity equation endowed with a source term $S'_{v}$: 
\begin{equation}
S_{v}=p\eta ^2\frac 1v\Phi _{h}\int_0^te^{-(1/2)nv\sigma _{tot}(t-t^{\prime
})}\cos (\eta (t-t^{\prime }))dt^{\prime }.  \label{B21}
\end{equation}
Considering $\mathbf{j}_{v}=-D\mathbf{\nabla }\Phi _{v}$ in order to take
into account the diffusion of the regenerated visible neutrons, we deduce
the expected diffusion equation in the steady state regime (i.e. $\partial
_tu_{v}=0$):
\begin{equation}
D\Delta \Phi _{v}=-S_{v}+\Sigma _a\Phi _{v}  \label{B20p}
\end{equation}
Thanks to the quickly oscillating terms related to $\cos (\eta t)$ in Eq. (%
\ref{B21}), and since the expression must be averaged against time assuming
a statistic distribution of the time origin $\varphi /\eta $ such that $%
\cos (\eta t+\varphi )$ ($\varphi $ is different for each neutron), we get: 
\begin{equation}
S_{v}\sim \frac{1}{2} \, p \, \Phi _{h} \, \Sigma _{s},  \label{sourceApp}
\end{equation}
where $\sigma_{tot}\approx\sigma_s$ for a regenerator with a low absorption cross section.

\bibliographystyle{unsrt}
\bibliography{Biblio3}

\end{document}